\documentstyle[12pt,epsfig,aaspp4]{article}

\begin{document}
\title{Neutrino Afterglows and Progenitors of Gamma-Ray Bursts}
\author{Z. G. Dai$^{1}$ and T. Lu$^{1,2}$}
\affil{$^1$Department of Astronomy, Nanjing University, Nanjing 210093, China}
\affil{$^2$IHEP, Chinese Academy of Sciences, Beijing 100039, China\\
E-mail: daizigao@public1.ptt.js.cn; tlu@netra.nju.edu.cn}

\begin{abstract}
Currently popular models for progenitors of gamma-ray bursts 
(GRBs) are the mergers of compact objects and the explosions of 
massive stars. These two cases have distinctive environments for GRBs: 
compact object mergers occur in the interstellar medium (ISM)  
and the explosions of massive stars occur in the preburst stellar wind. 
We here discuss neutrino afterglows from reverse shocks 
as a result of the interaction of relativistic fireballs with their 
surrounding wind matter. After comparing with the analytical result 
of Waxman \& Bahcall (2000) for the homogeneous ISM case,   
we find that the differential spectrum of neutrinos with energy 
from $\sim 3\times 10^{15}$ to $\sim 3\times 10^{17}$ eV in 
the wind case is softer by one power of the energy than in the ISM case. 
Furthermore, the expected flux of upward moving muons produced 
by neutrino interactions below a detector on the surface of 
the Earth in the wind case is $\sim 5$ events per year per km$^2$, 
which is about one order of magnitude larger than in the ISM 
case. In addition, these properties are independent of whether the fireballs 
are isotropic or beamed. Therefore, neutrino afterglows, if detected, may 
provide a way of distinguishing between GRB progenitor models based on 
the differential spectra of neutrinos and their event rates in a detector. 
\end{abstract}
 
\keywords{gamma-rays: bursts -- elementary particles 
-- shock waves -- stars: mass loss}

\section{Introduction}

The study of gamma-ray bursts (GRBs) has been 
revolutionized due to observations of multiwavelength 
afterglows in the past few years, but the nature of their progenitors 
remains unknown (for a review see M\'esz\'aros 1999). 
Two currently popular models for GRB progenitors are the  mergers 
of compact objects (neutron stars or black holes) and the 
explosions of massive stars. In the former model, compact objects 
are expected to have significant spacial velocities so that their 
mergers would take place at many kiloparsecs outside their birthplaces. 
Thus, GRBs produced by this model would occur in the 
interstellar medium (ISM) with density $n\sim 1\,{\rm cm}^{-3}$.   
Strong evidence for the massive star progenitor model has been recently
discovered. GRB 980425 was probably associated with the relatively
nearby Type Ic supernova (SN) 1998bw (Iwamoto et al. 1998; 
Kulkarni et al. 1998), and the supernova-like emission was also 
found in GRB 980326 (Bloom et al. 1999) and GRB 970228 
(Reichart 1999; Galama et al. 2000). These observations show that 
some or possibly all long-duration GRBs arise from the core collapse 
of massive stars. It has been widely believed that GRBs associated with 
supernovae should unavoidably occur in the preburst stellar wind environment 
with mass density $\rho \propto R^{-2}$. If GRB emission is isotropic, 
the X-ray and optical afterglow in the wind case must decline more rapidly 
than in the ISM case, as studied analytically by Dai \& Lu (1998), 
M\'esz\'aros, Rees \& Wijers (1998), Panaitescu, M\'esz\'aros 
\& Rees (1998) and Chevalier \& Li (1999, 2000). Guided by this 
argument, Dai \& Lu (1998) suggested for the first time that GRB 970616 
is a possible wind interactor based on the rapid fading indicated 
by two X-ray flux measurements. Recently Chevalier \& Li (1999) argued 
that GRB 980519 is an excellent wind interactor based on its X-ray, optical 
and radio data (although these observational data were shown analytically 
and numerically to be consistent with the dense medium model by Dai \& Lu 
[2000] and Wang, Dai \& Lu [2000]). Furthermore, the afterglow data
of some other bursts (e.g., GRB 970228, GRB 970508, GRB 980326 and 
GRB 980425) are also consistent with the wind environment model (Chevalier
\& Li 2000).  These afterglows have been argued to be further evidence 
for massive stars as GRB progenitors. 

In this paper we study neutrino afterglows from reverse shocks 
as a result of the interaction of relativistic fireballs with their 
surrounding wind matter by assuming that GRBs result from 
the explosions of massive stars.  We find that the differential spectrum 
of neutrinos below $\sim 3\times 10^{15}$ eV is proportional to 
$\epsilon_\nu^{-1}$ but the differential spectrum of neutrinos with energy 
from $\sim 3\times 10^{15}$ to $\sim 3\times 10^{17}$ eV steepens 
by one power of the energy. In addition, the expected flux of upward 
moving muons produced by neutrino interactions below a detector on the 
surface of the Earth is $\sim 5$ events per year per km$^2$ for typical 
parameters. We also find that this flux is $\sim 10$ times larger than estimated 
by Waxman \& Bahcall (2000), who studied neutrino emission from reverse 
shocks produced by the interaction of fireballs with the interstellar medium (ISM). 
 
The neutrinos are produced by $\pi^+$ created in interactions 
between accelerated protons and synchrotron photons from accelerated 
electrons in a relativistic fireball. This neutrino emission during 
the GRB phase was studied in the internal shock 
models by Waxman \& Bahcall (1997), Halzen (1998) and 
Rachen \& M\'esz\'aros (1998). It was found that a fraction, $\sim 0.1$, 
of the fireball energy would be converted by photomeson 
production to a burst of neutrinos with typical energy of a few $10^{14}$ eV 
(but also see Vietri 1998). The property of such neutrino bursts   
is independent of whether the ambient matter is a stellar 
wind or a constant density medium. Observations of these bursts
could be used to test the simultaneity of neutrino and photon
arrival to an accuracy of $\sim 1$ s, the weak equivalence 
principle, and the vacuum neutrino oscillation theory (Waxman 
\& Bahcall  1997).

The structure of this paper is as follows: In Section 2 we analyze reverse 
shocks produced during the interaction of ultra-relativistic fireballs with the 
surrounding wind matter and discuss the photon emission from these shocks. 
In Section 3 we investigate neutrino afterglow emission as a result of 
photo-meson interaction in the reverse shocks and in Section 4 
we discuss the detectability of such afterglows. In the final section, several
conclusions are drawn. 
 
\section{Shock Model and Photon Emission} 

We first assume that a relativistic GRB shell will interact
with the surrounding stellar wind via two shocks: a reverse shock 
and a forward shock. The forward shock runs forward into the wind
while the reverse shock sweeps up the shell material. The recently observed 
optical flash of GRB 990123 has been argued to come from 
a reverse shock (Akerlof et al. 1999; Sari \& Piran 1999; M\'esz\'aros \& 
Rees 1999). We believe that the reverse shock emission should be 
common for all GRBs. The shocked ambient and shell materials 
are in pressure balance and are separated by a contact discontinuity. 
We assume that these shocked materials are uniform and move together.
Sari \& Piran (1995) and Mitra (1998) considered the jump conditions 
for relativistic shocks and found the common Lorentz factor of 
the shocked materials measured in the unshocked medium frame,
\begin{equation} 
\gamma=\frac{\bar{\xi}^{1/4}\Gamma^{1/2}}{\sqrt{2}},
\end{equation}
where $\Gamma$ is the Lorentz factor of the unshocked shell measured 
in this frame and $\bar{\xi}\equiv \rho_{\rm sh}/\rho_{\rm w}$ is the ratio 
of proper mass densities of the unshocked shell and the unshocked ambient 
medium. The proper mass density of the ambient medium is expressed as 
\begin{equation}
\rho_w=\frac{\dot{M}_w}{4\pi R^2V_w}\equiv AR^{-2},
\end{equation}
where $\dot{M}_w$ and $V_w$ are the mass loss rate and wind velocity of 
the progenitor star, and $A\equiv\dot{M}_w/(4\pi V_w)=10^{-5}M_\odot\,
{\rm yr}^{-1}/(4\pi\times 10^3\,{\rm km}\,{\rm s}^{-1})A_*=5\times 10^{11}
\,{\rm g}\,{\rm cm}^{-1}A_*$ and $R$ is the radius of the shell in units of 1 cm  
(Chevalier \& Li 1999, 2000). The proper mass density of the unshocked shell 
is given by 
\begin{equation}
\rho_{\rm sh}=\frac{E_0}{4\pi R^2\Gamma^2c^2\Delta},
\end{equation}
where $E_0$ and $\Delta$ are the energy and the width (measured
in the unshocked medium frame) of the initial shell. A typical value 
of $A_*\sim 1$ for Wolf-Rayet stars is found from stellar mass-loss 
rates and wind velocities (Willis 1991; Chevalier \& Li 2000 ).
Since GRBs are believed to come from internal shocks,
$\Delta$ is approximately equal to the speed of light times the 
GRB durations and  thus its typical value should be $\sim 10$ light 
seconds. The rapid variability of GRBs and their nonthermal spectra 
require that $\Gamma$ be a few hundreds (Woods \& Loeb 1995). 
From the observed fluences of some GRBs and their measured redshifts,
$E_0$ is estimated to be between $10^{52}$ and $10^{54}$ ergs. 
A recent analysis by Freedman \& Waxman (2000) also gives this estimate. 
Scaling the involved quantities with these typical values,  
we find 
\begin{equation}
\bar{\xi}=655\frac{E_{53}}{A_*\Delta_{10}\Gamma_{300}^2}, 
\end{equation}
where $E_{53}=E_0/10^{53}{\rm ergs}$, 
$\Gamma_{300}=\Gamma/300$ and $\Delta_{10}$ is in units of
10 light seconds. From equations (1) and (4), the Lorentz factor of the shocked 
shell material is rewritten as
\begin{equation}
\gamma=62\frac{E_{53}^{1/4}}{A_*^{1/4}\Delta_{10}^{1/4}}.
\end{equation}
Following Sari \& Piran (1995) and Mitra (1998), we further derive 
the Lorentz factor of the shocked shell material measured 
in the unshocked shell rest frame,  
\begin{equation}
\gamma^\prime=\frac{\bar{\xi}^{-1/4}\Gamma^{1/2}}{\sqrt{2}}=
\frac{1}{2}\frac{\Gamma}{\gamma}=2.42\frac{A_*^{1/4}
\Delta_{10}^{1/4}\Gamma_{300}}{E_{53}^{1/4}},
\end{equation}
which implies that  the reverse shock is relativistic.  After the reverse shock 
passes through the shell, the shock front disappears. Instead of maintaining
a constant Lorentz factor (e.g., equation [6]), the shocked materials slow down 
with time based on the Blandford-McKee (1976) self-similar solution.
In the following we discuss photon emission from the reverse shock.

Because of pressure balance across the contact discontinuity, the shocked 
shell material and the shocked wind material have not only the same
bulk Lotentz factor but also the same internal energy density.
According to relativistic shock jump conditions, we obtain
the internal energy density of the shocked shell material,
\begin{equation}
e'=4\gamma^2\rho_{\rm w}c^2=2\Gamma\rho_{\rm w}c^2\bar{\xi}^{1/2}. 
\end{equation}
We assume that $\epsilon_e$ and $\epsilon_B$ are 
the fractions of the internal energy density (in the shocked material rest frame) 
that are carried by electrons and magnetic fields respectively. The minimum 
Lorentz factor of the electrons accelerated behind the reverse shock is 
approximated by $\gamma_m\approx (m_p/m_e)\epsilon_e\gamma^\prime$ 
(Waxman \& Bahcall 2000), viz., 
\begin{equation}
\gamma_m\approx 445\epsilon_{e,-1}
\left(\frac{A_*^{1/4}\Delta_{10}^{1/4}\Gamma_{300}}{E_{53}^{1/4}}\right), 
\end{equation}
where $\epsilon_{e,-1}=\epsilon_e/0.1$.
Moreover, the magnetic field strength in the shocked shell material is given by 
\begin{equation}
B'=(8\pi \epsilon_B e')^{1/2}
    =\frac{(16\pi\epsilon_B\Gamma A\bar{\xi}^{1/2})^{1/2}}{t_b},
\end{equation}
where $t_b=R/c$ is the time in the burster's rest frame. Substituting the relation 
between this time and the observed time ($t$), $t_b=2\gamma^2 t /(1+z)$, into 
the above equation, and using equations (4) and (5), we further have 
\begin{equation}
B'=3.8\times 10^3\epsilon_{B,-3}^{1/2}\left(\frac{1+z}{2}\right)
\left(\frac{\Delta_{10}^{1/4}A_*^{3/4}}{E_{53}^{1/4}t_s}
\right)\,{\rm G},
\end{equation}
where $\epsilon_{B,-3}=\epsilon_B/10^{-3}$, $z$ is the redshift
of the source and $t_s=t/ 1\,{\rm s}$. It should be noted  
that $\epsilon_e\sim 0.1$ (Freedman \& Waxman 2000), but $\epsilon_B$
is highly uncertain and its reasonable value may be taken from $\sim 10^{-2}$ 
to $\sim 10^{-6}$. Several previous studies of GRB afterglows (e.g., Galama 
et al. 1999; Dai \& Lu 1999, 2000; Wang, Dai \& Lu 2000) gives $\epsilon_B
\sim 10^{-4}-10^{-6}$. A recent detailed study of the afterglows of GRBs 
980703, 990123 and 990510 by Panaitescu \& Kumar (2000) leads to 
$\epsilon_B\sim 10^{-3}-10^{-4}$. In addition, Holland et al. (2000) find
that $\epsilon_B$ is as small as $10^{-5}$. Therefore, we choose a typical 
value: $\epsilon_B\sim10^{-3}$. 
 
Let's consider synchrotron radiation of the electrons accelerated behind 
the reverse shock. We first derive two characteristic frequencies of 
synchrotron photons:  the typical frequency $\nu_m$ corresponding to 
the minimum electron Lorentz factor and the cooling frequency $\nu_c$. 
From equations (5), (8) and (10), we obtain the typical frequency 
in the observer frame, 
\begin{equation}
\nu_m = \frac{\gamma\gamma_m^2}{1+z}
                 \frac{eB'}{2\pi m_ec}
         =6.3\times 10^{16}\epsilon_{e,-1}^2\epsilon_{B,-3}^{1/2}
         \frac{A_*\Delta_{10}^{1/2}\Gamma_{300}^2}
         {E_{53}^{1/2}t_s}\,\,{\rm Hz}
\end{equation}
The cooling frequency corresponds to the cooling Lorentz factor $\gamma_c$, 
at which an electron is cooling on the dynamical expansion time.
We believe that this Lorentz factor in the reverse shock will increase with time
because of the cooling timescale $\propto B'^{-2}\propto t^2$. Initially, 
$\gamma_c-1\le 1$. At this stage, a cooling electron may be non-relativistic 
and its kinetic energy $E_e\approx (1/2)\gamma m_ec^2\beta^2$. Thus, its cooling 
timescale due to cyclotron radiation, measured in the observer's frame, can be 
estimated as $t_0=(1+z)E_e/P_{\rm cyc}(\beta)$, where $P_{\rm cyc}(\beta)
=(4/3)\sigma_Tc\gamma^2\beta^2B'^2/(8\pi)$ is the cyclotron radiation power
(in the local observer frame) of an electron with velocity of $\beta c$ and with 
$\sigma_T$ being the Thomson scattering cross section. Using equation (10),
we easily find 
\begin{equation} 
t_0=1.0\epsilon_{B,-3}\left(\frac{1+z}{2}\right)\frac{\Delta_{10}^{1/4}
A_*^{5/4}}{E_{53}^{1/4}}\,{\rm s}.
\end{equation}
Please note that $t_0$ is independent of $\beta$. This implies that at $t<t_0$ 
an electron accelerated to $\gamma_m$ in the magnetic field $B'$ will be 
able to cool to become non-relativistic, initially through synchrotron radiation 
and subseqently through cyclotron radiation, on the dynamical expansion 
time $t$. However, when $t>t_0$, the magnetic field ($B'\propto t^{-1}$) 
will become weaker and the cooling timescale due to cyclotron radiation 
must be longer than $t$, so an electron with $\gamma_m$ cannot cool 
to a non-relativistic velocity on time $t$. In this case, cyclotron radiation 
is no longer a cooling mechanism but it should be replaced 
by synchrotron radiation. Since for typical parameters $t_0\sim 1$ s,
which is much less than the durations of long GRBs from the collapse 
of massive stars, we will discuss the photon and neutrino emission from 
the reversely shocked matter at $t>t_0$ in the remaining text. According 
to Sari, Piran \& Narayan (1998), the cooling Lorentz factor is defined by
\begin{equation}
\gamma\gamma_cm_ec^2=P_{\rm syn}(\gamma_c)t/(1+z),
\end{equation}
where $P_{\rm syn}(\gamma_c)=(4/3)\sigma_Tc\gamma^2\gamma_c^2
\beta^2B'^2/(8\pi)$ is the synchrotron radiation power (in the local observer 
frame) of an electron with Lorentz factor of $\gamma_c$ (and $\beta\sim 1$).  
Equation (13) leads to 
\begin{equation}
\gamma_c=\frac{6\pi m_ec(1+z)}{\sigma_T\gamma B'^2t}
=2.0\epsilon_{B,-3}^{-1}\left(\frac{1+z}{2}\right)^{-1}
\frac{E_{53}^{1/4}t_s}{\Delta_{10}^{1/4}A_*^{5/4}}.
\end{equation}
It is clear that $\gamma_c\gg 1$ for $t\gg 1$ s, implying the cooling 
electrons are indeed relativistic. Using this equation, we further derive 
the cooling frequency in the observer frame,
\begin{equation}
\nu_c= \frac{\gamma\gamma_c^2}{1+z}
                 \frac{eB'}{2\pi m_ec}
         =1.3\times 10^{12}\epsilon_{B,-3}^{-3/2}
           \left(\frac{1+z}{2}\right)^{-2}
         \frac{E_{53}^{1/2}t_s}{A_*^2\Delta_{10}^{1/2}}\,\,{\rm Hz}.
\end{equation}
We can see from equations (11) and (15) that for typical parameters  
the cooling frequency is much lower than the typical frequency,
indicating that all radiating electrons cool rapidly down to the cooling 
Lorentz factor. In other words, the shocked shell material is in the fast 
cooling regime. It is interesting to note that this conclusion has also been 
drawn by Chevalier \& Li (2000). Therefore, the observed specific luminosity 
peaks at $\epsilon_c\equiv h\nu_c$ rather than $\epsilon_m\equiv h\nu_m$, 
with a peak value approximated by
\begin{equation}
L_{\epsilon_c}=(2\pi\hbar)^{-1} (1+z)\gamma N_eP'_c
=2.7\times 10^{62}\epsilon_{B,-3}^{1/2}
  \left(\frac{1+z}{2}\right)\frac{E_{53}A_*^{1/2}}
  {\Gamma_{300}\Delta_{10}}\,\,{\rm s}^{-1},
\end{equation}
where  $N_e=E_0ct/[(1+z)\Gamma m_pc^2\Delta]$ is
the number of radiating electrons in the shocked shell region,
and $P'_c=m_ec^2\sigma_TB'/(3e)$ is the power radiated 
per electron per unit frequency in the shocked shell rest frame.  

We turn to derive the synchrotron self-absorption frequency of the
reversely shocked matter. In the comoving frame of the shocked matter,
the absorption coefficient for $\nu'>\nu_c'$ is given by
\begin{equation}
\alpha'_{\nu'}=\frac{\sqrt{3}e^3}{8\pi m_e}\left(\frac{3e}{2\pi m_e^3c^5}
\right)^{p/2}(m_ec^2)^{p-1}K\lambda B'^{(p+2)/2}\nu'^{-(p+4)/2}
\Gamma\left(\frac{3p+2}{12}\right)\Gamma\left(\frac{3p+22}{12}\right),
\end{equation}
where $K=4(p-1)\gamma'\gamma_c(\rho_{\rm sh}/m_p)$ and $\lambda=(1/2)
\int^\pi_0(\sin\alpha)^{(p+2)/2}\sin\alpha d\alpha$ (Rybicki \& Lightman 1979). 
In the present case, the electron distribution index $p=2$ because $\nu_c\ll\nu_m$, 
and the width of the reverse shock $R'=\gamma ct/(1+z)$.  Setting $\tau(\nu'_a)
\equiv\alpha'_{\nu'_a}R'=1$, we can derive the synchrotron self-absorption 
frequency in the obsever's frame,
\begin{equation}
\nu_a=\frac{\gamma\nu'_a}{1+z}=2.2\times 10^{15}\left(\frac{1+z}{2}\right)
^{-1/3}\frac{A_*^{1/6}E_{53}^{1/6}}{\Delta_{10}^{1/6}\Gamma_{300}^{1/3}
t_s^{2/3}}\,\,{\rm Hz}.
\end{equation}
Hence, $\nu_c\ll \nu_a\ll \nu_m$ for typical parameters. 

We assume that the electrons behind the reverse shock follow a power law 
energy distribution, $dn'_e/d\gamma_e\propto\gamma_e^{-2}$ for 
$\gamma_e \ge \gamma_m$ (Blandford \& Eichler 1987). In this case, 
the synchrotron radiation spectrum is a broken power law: 
\begin{equation}
L_{\epsilon_\gamma}=\left\{
       \begin{array}{lll}
           L_{\epsilon_c}(\epsilon_a/\epsilon_c)^{-1/2}(\epsilon_\gamma/\epsilon_a)
              ^{5/2} & {\rm if}\,\,\,\,\epsilon_\gamma\le\epsilon_a \\
           L_{\epsilon_c}(\epsilon_\gamma/\epsilon_c)^{-1/2} & {\rm if}\,\,\,\,
                   \epsilon_a\le\epsilon_\gamma\le\epsilon_m \\
           L_{\epsilon_c}(\epsilon_m/\epsilon_c)^{-1/2}
                  (\epsilon_\gamma/\epsilon_m)^{-1} & {\rm if}\,\,\,\,
                  \epsilon_\gamma\ge\epsilon_m,
       \end{array}
       \right.
\end{equation}
where $\epsilon_a=h\nu_a$. The protons behind the reverse shock are expected 
to be accelerated to the same power-law distribution as the electrons (with 
the maximum proton energy which will be estimated in the next section).

\section{Neutrino Emission}

For convenience, we hereafter denote the particle energy measured 
in the shocked shell rest frame with a prime, and the particle 
energy in the observer frame without prime, e.g., 
$\epsilon_\gamma=\gamma \epsilon_\gamma'/(1+z)$.   
We now consider neutrino production in the wind case. Assuming 
$n'_\gamma(\epsilon'_\gamma)d\epsilon'_\gamma$ to be 
the photon number density in the shocked shell rest frame and 
following Waxman \& Bahcall (1997), we can write the fractional 
energy loss rate of a proton with energy $\epsilon'_p$ 
due to pion production,
\begin{equation}
t_\pi^{'-1}(\epsilon'_p)\equiv -\frac{1}{\epsilon'_p}
\frac{d\epsilon'_p}{dt'}
=\frac{1}{2\gamma_p^2}c\int^\infty_{\epsilon_0}
d\epsilon\sigma_\pi(\epsilon)\xi(\epsilon)\epsilon
\int^\infty_{\epsilon/2\gamma_p} dxx^{-2}n'(x),
\end{equation}
where $\gamma_p=\epsilon'_p/m_pc^2$, $\sigma_\pi(\epsilon)$
is the cross section for pion production for a photon with energy 
$\epsilon$ in the proton  rest frame, $\xi(\epsilon)$ is the average
fraction of energy lost to the pion, $\epsilon_0=0.15$ GeV is the 
threshold energy, and the photon number density is related to the observed
specific luminosity by $n'(x)=L_{\epsilon_\gamma}(\gamma x)/[4\pi R^2c
(1+z)\gamma x]$. Because of the $\Delta$ resonance, we find that
photo-meson production is dominated by the interaction with photons
in the energy range $\epsilon_\gamma >\epsilon_m$. Considering the 
photon spectrum in this energy range, equation (20) leads to
\begin{equation}
 t_\pi^{'-1}(\epsilon'_p)=\frac{L_{\epsilon_c}}{3\pi(1+z)R^2
\gamma}\left(\frac{\epsilon'_c}{\epsilon'_m}\right)^{1/2}
\left(\frac{\gamma_p\epsilon'_m}{\epsilon_{\rm peak}}\right)
\frac{\sigma_{\rm peak}\xi_{\rm peak}\Delta\epsilon}
{\epsilon_{\rm peak}},
\end{equation}
where $\sigma_{\rm peak}=5\times 10^{-28}\,{\rm cm}^{-2}$ and
$\xi_{\rm peak}=0.2$ at the resonance $\epsilon=\epsilon_{\rm peak}
=0.3$ GeV, and $\Delta\epsilon=0.2$ GeV is the peak width.
The fraction of energy loss of
a proton with observed energy $\epsilon_p$ by pion production, 
$f_\pi (\epsilon_p)$, is defined by $t_{\pi}^{'-1}$ times the expansion 
time of the shocked shell material ($\sim R/\gamma c$). 
Thus, we have   
\begin{equation}
f_\pi (\epsilon_p) = 2.0\epsilon_{e,-1}\left(\frac{1+z}{2}
\right)^{-2}\left(\frac{A_*^{3/2}\Delta_{10}^{1/2}}{E_{53}^{1/2}}
\right)\left(\frac{\epsilon_p}{10^{17}{\rm eV}}\right).
\end{equation}
It is interesting to note that $f_\pi (\epsilon_p)$ is independent of $\epsilon_B$ 
and $\Gamma$. Similarly  to the cooling electron Lorentz factor defined by 
Sari et al. (1998) (see equation [13]), we can define the cooling proton energy 
$\epsilon_{p,c}$ based on $f_\pi(\epsilon_{p,c})=1$. According to equation 
(22), we find $\epsilon_{p,c}\approx 5\times 10^{16}$ eV for typical 
parameters. This implies that the protons with energy 
$\ge \epsilon_{p,c}$ accelerated behind the reverse shock must 
lose almost all of their energy (viz., significant cooling) due to 
photo-meson interactions, but the protons with energy 
$<\epsilon_{p,c}$ lose only a fraction ($\sim f_\pi$) of their energy. 

We now turn to discuss the neutrino spectrum. 
The photo-meson interactions include (1) production of $\pi$ mesons:
$p\gamma\rightarrow p+\pi^0$ and $p\gamma\rightarrow n+\pi^+$,
and (2) decay of $\pi$ mesons: $\pi^0\rightarrow 2\gamma$ and
$\pi^+\rightarrow \mu^++\nu_\mu\rightarrow e^++\nu_e+
\bar{\nu}_\mu+\nu_\mu$. These processes produce neutrinos with energy 
$\sim 0.05\epsilon_p$ (Waxman \& Bahcall 1997). Since the protons 
with energy $<\epsilon_{p,c}$ lose only a fraction ($\sim f_\pi\propto 
\epsilon_p$) of their energy,  the differential spectrum of neutrinos below 
the break energy $\sim 3\times 10^{15}$ eV is harder than the proton 
spectrum by one power of the energy. But since the protons with energy 
$\ge \epsilon_{p,c}$ accelerated behind the reverse shock must 
lose almost all of their energy, the neutrino spectrum above the break 
traces the proton spectrum. Therefore, if the differential spectrum of 
accelerated protons is assumed to be a power law form $n(\epsilon_p)
\propto\epsilon_p^{-2}$,  the differential neutrino spectrum is 
$n(\epsilon_\nu)\propto \epsilon_\nu^{-1}$ below the break and 
$n(\epsilon_\nu)\propto \epsilon_\nu^{-2}$ above the break.  

The maximum energy of the resultant neutrinos is estimated 
as follows. This energy is determined by the maximum energy 
of  the protons accelerated by the reverse shock. The typical Fermi 
acceleration time is $t'_a=fR_L/c$, where $R_L=(1+z)\epsilon_p/
(\gamma eB')$ is the Larmor radius and $f$ is of order unity (Hillas 1984). 
The requirement that this acceleration time is equal to the time for energy 
loss of protons ($t'_\pi$) due to pion production leads to the maximum 
proton energy,
\begin{equation}
\epsilon_{p,{\rm max}}=5.6\times 10^{18}
f^{-1/2}\epsilon_{e,-1}^{-1/2}\epsilon_{B,-3}^{1/4}
\left(\frac{1+z}{2}\right)^{1/2}\frac{E_{53}^{3/8}}
{A_*^{5/8}\Delta_{10}^{3/8}}\,\,{\rm eV}.
\end{equation}
From this equation, we can draw two conclusions: (1) For reasonable 
parameters, the maximum proton energy is $\sim 6\times 10^{18}$ eV, 
which is two orders of magnitude smaller than the maximum energy 
of the protons accelerated by the reverse shock in the ISM case 
(Waxman \& Bahcall 2000). The physical conditions in the reverse shock 
for the ISM case imply that protons can be Fermi accelerated 
to $\sim 10^{21}$ eV (Waxman 1995; Vietri 1995; Milgrom \& Usov 
1995; see Waxman [1999, 2000] for recent reviews). (2) The maximum energy 
of neutrinos produced in the wind case is $\sim 3\times 10^{17}$ eV.  

\section{Detectability}

We discuss the detectability of the neutrino afterglow 
emission in the wind case. Since  the protons with energy 
$\ge 5\times 10^{16}$ eV must lose almost all of 
their energy due to photo-meson interactions, the present day 
neutrino energy density due to GRBs is approximately given
by $U_\nu=(1/2)(1/2)t_H\dot{E}$, where the first factor $1/2$ accounts for 
the fact that about one half of the proton energy is lost to neutral pions 
which do not produce neutrinos, the second factor $1/2$ accounts for 
the fact that about one half of the energy in charged pions is transferred to 
$\nu_\mu+{\bar{\nu}}_\mu$, and $t_H=10$ Gyr is the Hubble time. 
Here we assume that  $\dot{E}=0.8\times 10^{44}{\rm erg}\,
{\rm Mpc}^{-3}\,{\rm yr}^{-1}$ is the production rate of GRB energy 
per unit volume (Waxman \& Bahcall 2000). The neutrino 
flux is thus approximated by
\begin{equation}
\phi_\nu \approx \frac{c}{4\pi}\frac{U_\nu}{\epsilon_\nu} 
\approx 4\times 10^{-15}\left(\frac{\epsilon_\nu}
{3\times 10^{15}{\rm eV}}\right)^{-1}\,\,{\rm cm}^{-2}
\,{\rm s}^{-1}\,{\rm sr}^{-1}.
\end{equation} 
The resulting high-energy neutrinos may be observed by detecting
the Cherenkov light emitted by upward moving muons produced
by neutrino interactions below a detector on the surface of the
Earth (Gaisser, Halzen \& Stanev 1995; Gandhi et al. 1998). 
Planned 1 km$^3$ detectors of high energy neutrinos include 
ICECUBE, ANTARES, NESTOR (Halzen 1999) and NuBE
(Roy, Crawford \& Trattner 1999).
The probability that a neutrino could produce a high-energy muon 
in the detector is approximated by $P_{\nu\rightarrow \mu}\approx 
6\times 10^{-4}(\epsilon_\nu/3\times 10^{15}{\rm eV})^{0.5}$. 
Using equation (24), we obtain the observed neutrino event rate 
in a detector, 
\begin{equation}
N_{\rm events}=2\pi \phi_\nu P_{\nu\rightarrow \mu}
\approx 5 \left(\frac{\epsilon_\nu}{3\times 10^{15}
{\rm eV}}\right)^{-0.5}\,\,{\rm km}^{-2}\,{\rm yr}^{-1}.
\end{equation}
This equation shows that a km$^2$ neutrino detector should 
detect each year about 5 neutrinos (with energy of
$\sim 3\times 10^{15}$ eV) correlated with GRBs. For a GRB, 
its neutrino emission from the reverse shock in the wind case
should be delayed to a few seconds after the main burst. 
Waxman \& Bahcall (2000) have found $f_\pi \sim 0.1$ 
for neutrino emission from reverse shocks in the ISM case
(where the typical energy of neutrinos is $\sim 3\times 10^{17}$ eV). 
Using the same expression of $P_{\nu\rightarrow\mu}$,  
we have re-derived their neutrino event rate in a detector and 
obtained $N_{\rm events}\sim 0.5\,{\rm km}^{-2}\,{\rm yr}^{-1}$,
which is smaller than our event rate by a factor of $\sim 10$.

\section{Discussion and Conclusions}
  
Neutrino bursts during the GRB phase were studied in the  internal shock 
models by Waxman \& Bahcall (1997) and Halzen (1998) who found that the 
neutrino event rate in a detector (mainly neutrinos with typical energy of 
a few $10^{14}$ eV) is $\sim 26$ events per year per km$^2$, which
is larger than our event rate by a factor of $\sim 5$.  
Compared with the analytical result of 
Waxman \& Bahcall (2000), our discussions on neutrino afterglows in the 
wind case can lead to the following conclusions: (1) The protons with energy 
$\ge 5\times 10^{16}$ eV must lose almost all of their energy due to 
photo-meson interactions and thus the neutrino afterglow emission in the wind 
case is dominated by neutrinos with energy $\sim 3\times 10^{15}$ eV. 
(2) The maximum energy of the protons accelerated behind the reverse 
shock in the wind case is $\sim 6\times 10^{18}$ eV, so ultrahigh energy 
cosmic rays cannot be produced in this case.  In addition, the maximum 
neutrino energy is $\sim 3\times 10^{17}$ eV. (3) The neutrino differential 
spectrum below $\sim 3\times 10^{15}$ eV is proportional to 
$\epsilon_\nu^{-1}$ but the spectrum between $\sim 3\times 10^{15}$ 
and $\sim 3\times 10^{17}$ eV steepens by one power of the energy. 
(4) The observed neutrino event rate in the wind case is $\sim 10$ times 
larger than the one in the ISM case. 

If GRB emission is isotropic, the optical afterglow in the wind case must 
decline more steeply than in the ISM case. This has been suggested as 
a plausible way of distinguishing between the GRB progenitor models   
(Chevalier \& Li 1999, 2000). It is widely believed that GRBs may come 
from jets (Kulkarni et al. 1999; Castro-Tirado et al. 1999). As argued by
Rhoads (1999) and Sari, Piran \& Halpern (1999), the optical afterglow
from a jet is likely to decay more rapidly at late times than 
at the early stage due to the lateral spreading effect. If this effect is true, 
however, both ISM and wind cases should show the same emission feature 
during the lateral spreading phase, and in particular on a timescale
of days the wind density is similar to typical ISM densities so that 
an interaction with the wind would give results that are not too different
from the ISM case (Chevalier \& Li 2000; Livio \& Waxman 1999). If GRBs 
are beamed, thus, their optical afterglow emission could not be used 
to discriminate the massive star progenitor model from the compact binary 
progenitor model. However, the neutrino afterglow emission discussed here is
independent of whether the fireballs are isotropic or highly collimated. 
Therefore, neutrino afterglows, if detected, may be used to distinguish 
between GRB progenitor models based on differential spectra of observed 
neutrinos and their event rates in a detector. 

What we want to point out is that the above conclusions are drawn by 
considering typical values of the parameters entering the fireball shock model. 
In fact, these parameters may have respective distributions. It is interesting 
to note that such distributions may lead to an event rate larger than 
estimated in this paper (Halzen \& Hooper 1999).

\acknowledgments

We would like to thank Prof. F. Halzen and the referee for their valuable 
comments that enabled us to improve our paper, Y. F. Huang 
for reading carefully this paper, and X. Y. Wang for useful discussions. 
This work was supported partially by the National Natural Science 
Foundation of China (grants 19825109 and 19773007), and  
by the National Project on Fundamental Researches.


\end{document}